\documentclass[twocolumn,twoside,slac_two]{revtex4}

\usepackage[T1]{fontenc}
\usepackage[ansinew]{inputenc}
\usepackage[scaled]{helvet}
\usepackage{amsmath, amsfonts, amssymb, bm, amsthm} 
\usepackage{mathrsfs}
\usepackage{fancyhdr}
\usepackage{graphicx,textcomp,booktabs,amsmath}
\usepackage{txfonts}
\usepackage{subfigure}
\usepackage{natbib}

\def\bfield{B = \left(124\,{}\pm6\,\mathrm{(stat.)}\,{}^{+15}_{-6}\,\mathrm{(sys.)}\right)\,\mu\mathrm{G}}

\newcommand{\Diff}[2]{\frac{\text{d} #1}{\text{d} #2}}

\newcommand{\Difft}[2]{\text{d} #1 \slash \text{d} #2}

\newcommand{\unit}[1]{\,\text{#1}}

\newcommand{\Nel}{N_\text{el}}

\begin{document}

\title{Cross Calibration of Imaging Air Cherenkov Telescopes with Fermi}

%

\author{Manuel Meyer{\email{manuel.meyer@desy.de}},  Hannes-Sebastian Zechlin,  Dieter Horns}
\affiliation{Institut f\"ur Experimentalphysik, University of Hamburg\\Luruper Chaussee 149 D-22767, Hamburg, Germany}
   \date{ \today}

\begin{abstract}
An updated model for the synchrotron and inverse Compton emission from a population of high energy electrons of the Crab nebula is used to reproduce the measured spectral energy distribution from radio to high energy $\gamma$-rays.  By comparing the predicted inverse Compton component with recent Fermi measurements of the nebula's emission, it is possible to determine the average magnetic field in the nebula and to derive the underlying electron energy distribution. The model calculation can then be used to cross calibrate the Fermi observations with ground based air shower measurements. The resulting energy calibration factors are derived and can be used for combining broad energy measurements taken with Fermi in conjunction with ground based measurements.

\end{abstract}

\maketitle

\thispagestyle{fancy}


\section{Introduction}

The Crab Nebula is probably the best studied object in astrophysics (for a
recent review see e.g. \citet{2008ARA&A..46..127H}). It is the remnant of a
core-collapse supernova which occured in 1054 AD at a distance of $d \approx 2
\unit{kpc}$ \citep{1968AJ.....73..535T}. Near its geometric center resides
a pulsar which continuously injects a wind of ultra-relativistic particles into
the nebula.  The wind terminates at a shock front where electrons are pitch
angle isotropized, forming a broad power-law in energy.  

Observations of the synchrotron and inverse Compton nebula have been carried
out in every accessible wavelength resulting in a remarkably well known
spectral energy distribution (SED). The large area telescope (LAT) onboard the Fermi satellite
\citep{2009arXiv0911.2412T} has recently measured the flux of the $\gamma$-rays
of the Crab between $100\unit{MeV}$ and $300\unit{GeV}$ with unpreceded
accuracy.

These observations can be used for a cross calibration between the Fermi/LAT and ground based air shower experiments.
This leads to the elimination of the systematic uncertainties on the absolute energy scale of typically 15~\% for imaging air
Cherenkov telescopes (IACTs). 

For the cross calibration we use our approach presented in \citet{Meyer}:   
based on the work of \citet{1998ApJ...503..744H},
we extract the distribution of electrons from the synchrotron
spectrum under the assumption of a constant magnetic field strength throughout
the nebula.  Using this electron distribution in conjunction with the seed
photon fields as extracted from observations, we obtain a detailed prediction for the
inverse Compton emission. 

The average magnetic field inside the nebula is determined by fixing the model to the Fermi/LAT observations.
The model can directly be used as the basis for the cross calibration.  For this purpose
energy scaling factors are derived to correct the measurements of the ground
based instruments to the model. 

The article is organized as follows. In section \ref{sec:SED} the model for the SED is quickly reviewed, where the details 
and the discussion of the underlying electron spectrum are spared out and we refer the reader to \citet{Meyer}. The results of the 
cross calibration are presented in section \ref{sec:crosscal} together with an
application to extract limits on the  diffuse $\gamma$-ray background at TeV
energies. Section \ref{sec:summary} summarizes the article and gives a short
outlook on possible future work.

\section{Model of the Spectral Energy distribution of the Crab Nebula}
\label{sec:SED}

A population of relativistic electrons is assumed to be distributed in a
spherical volume following a Gaussian density distribution with its maximum at the nebula's center.  The width of the
Gaussian is parameterized in order to reproduce the shrinking size of the
nebula with increasing frequency. The volume of the nebula is assumed to be
filled with an entangled magnetic field of constant field strength. Within the
volume occupied by the electrons, various seed photons are upscattered.  The
effective density of seed photons $n_\mathrm{seed}$ is simply found by
convolving the electron density with the
photon density \citep[see][for further details]{1998ApJ...503..744H}.
The total emission from the nebula is then found by integration:
\begin{eqnarray}
 L_\nu &=& \int~\mathrm{d}\gamma~ n(\gamma) \left (\mathcal{L}_\nu^\mathrm{Sy} + 
\mathcal{L}_\nu^\mathrm{IC}\right),
\end{eqnarray}
with $\mathcal{L}_\nu$ the single particle emission functions for
synchrotron (Sy) and inverse Compton processes (IC) \citep[see e.g.][]{1970RvMP...42..237B}:
\begin{eqnarray}
 \mathcal{L}_\nu^\mathrm{Sy} &=& \frac{\sqrt 2 e^3 B }{m c^2}~\frac{\nu}{\nu_\mathrm{c}}\int_{\nu\slash\nu_\mathrm{c}}^{~\infty} K_{5\slash 3} (x)~ \mathrm d x\, , \\ 
 \mathcal{L}_\nu^\mathrm{IC} &=& \frac 3 4 ~\frac{\sigma_\mathrm{T}c}{\gamma^2}h\nu\int_{h\nu\slash(4\gamma^2)}^{~h\nu}~\mathrm d \epsilon ~\frac{n_\mathrm{seed}(\epsilon)}{\epsilon}f_\mathrm{IC}(\epsilon,\nu,\gamma),\label{eqn:ic} 
\end{eqnarray}
where we have averaged the pitch angle to give $\sqrt{2\slash3}$ and $\sigma_\mathrm{T}$ denotes the Thomson cross-section. The critical frequency $\nu_\mathrm{c}$ is defined as 
\begin{eqnarray}
\nu_\mathrm{c} = \frac{3 e}{4\pi m c}~B~\gamma^2,
\end{eqnarray}
and $K_{5\slash3}(x)$ stands for the modified Bessel function of fractional order $5\slash3$. Introducing the kinematic variable $q$,
\begin{eqnarray}
q = \frac{h \nu}{4\epsilon\gamma^2[1-h\nu\slash(\gamma m c^2)]},
\end{eqnarray}
the IC distribution function $f_\mathrm{IC}$ can be written as
\begin{eqnarray}
f_\mathrm{IC}(\epsilon,\nu,\gamma) & = & \nonumber\\
  2q\ln q & + & (1+2q)(1-q) \nonumber\\
  &{}&+ \frac 1 2~\frac{\left[{4\epsilon\gamma q}/({mc^2})\right]^2}{1+{4\epsilon\gamma q^2}/{(mc^2)}}(1-q).
\end{eqnarray}

\begin{figure*}[t!]
 \centering
 \includegraphics[width = 0.9\textwidth]{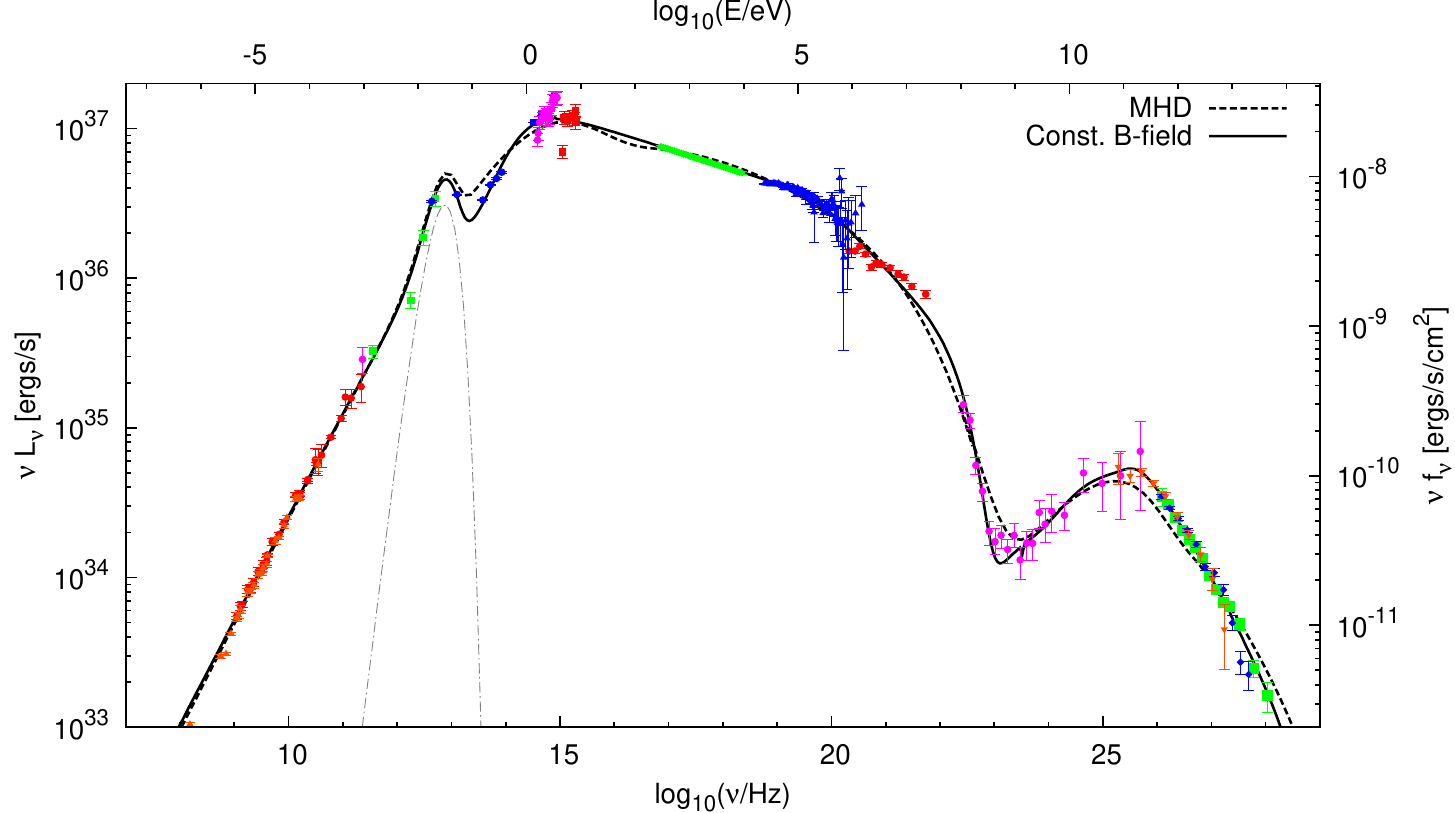}
 \caption{Broadband SED of the Crab nebula. See section \ref{sec:SED} for details. The two black curves correspond to the models described in \citet{Meyer}}
 \label{fig:SED}
\end{figure*}

For the inverse Compton channel photons from several seed photon fields
are taken into account: (1) synchrotron radiation, (2) emission from thermal
dust, (3) the cosmic microwave background (CMB), and (4) optical line emission
from the nebula's filaments. The seed photon density $n_\mathrm{seed}$ in Eqn. \ref{eqn:ic} is the sum of all these components. 
Like the electron population, the spatial photon densities are
approximated by Gaussian distributions whereas the photon density of the CMB is
taken to be constant throughout the nebula. The spatial variance of these
distributions is also energy dependent. The resulting broadband energy
distribution is shown in Fig. \ref{fig:SED}. The electron spectrum is varied
until the resulting synchrotron spectrum reproduces the observational data.
\\
The compilation of data used here is summarized  
in \citet{2004ApJ...614..897A} and references therein. Additionally,
new data are added which are listed in  Table \ref{tbl:data}. The solid black
curve in Fig. \ref{fig:SED} is the sum of all contributions including
synchrotron and IC emission as well as thermal emission from dust in the nebula
and optical line emission from the filaments. For the thermal dust emission a
graybody spectrum is used and a temperature of $T = 93 \unit{K}$ was derived by
fitting  the combined spectrum (thermal and non-thermal emission) to the data (solid gray line in Fig. \ref{fig:SED}).
The flux of the line emissions (orange solid line) is taken from
\citet{1985ARA&A..23..119D,1987AJ.....94..964D} and
\citet{1990ApJ...357..539H}.  The optical line emission of the filaments in the nebula 
is estimated in the following way:  The high resolution spectral observations of individual filaments have been corrected for 
extinction \citep{1990ApJ...357..539H} and scaled to match the global emission from the filaments 
 \citep[see e.g. the discussion in][]{1985ARA&A..23..119D}. \\
The FIR observations from Spitzer, ISO,  and Scuba (orange and magenta circles
 in Fig.~\ref{fig:SED}respectively) deviate from the power-law extrapolation
of the radio spectra.  In the framework of two distinct electron populations,
the shape of the continuum is naturally explained by the transition of
the two synchrotron emission components. 
\\
 The dashed blue curve indicates the total synchrotron and IC emission which
results from the contributions from wind electrons whereas the dashed red line
shows the contribution of the radio electrons. The dashed cyan lines show the total 
synchrotron and IC flux from both populations, respectively. 

\begin{table}[hbt]
\centering
 \begin{small}
\begin{tabular}{lll}
\textbf{Energy Band} & \textbf{Instrument} & \textbf{Reference}\\
\hline
\hline
Sub milimeter		& ISO \& SCUBA			&  {\citet{2004MNRAS.355.1315G}}  \\
 to far infrared	& SPITZER			&  {\citet{2006AJ....132.1610T}} \\
 \hline 
 X-Ray to 		& XMM-Newton			& {\citet{2005SPIE.5898...22K}}\\
 $\gamma$-rays		& SPI				& {\citet{2009ApJ...704...17J}}\\
 {}			& IBIS$\slash$ISGRI 		& {\citet{2008int..workE.144J}}\\
 {}			& Fermi / LAT			& {\citet{2009arXiv0911.2412T}}\\
 \hline
 VHE			& H.E.S.S.			& {\citet{2006A&A...457..899A}}\\
 {}			& MAGIC				& {\citet{2008ApJ...674.1037A}}
\end{tabular}
\end{small}
\caption[Observations for SED of the Crab]{References for the observations used for the SED. All other data are taken from \citet{2004ApJ...614..897A} and references therein.}
\label{tbl:data}
\end{table}

The electron spectrum was adapted such that the resulting synchrotron emission
reproduces the power law measured with XMM-Newton. Both, the XMM-Newton and
INTEGRAL (with the instruments SPI and IBIS/ISGRI) observatories are calibrated
on the basis of detailed simulation and laboratory measurements. This approach
differs from widely used corrections of the response function in order to
re-produce a specific spectral shape and flux of the Crab nebula. The
corrected measurements are therefore model-dependent and have not been 
included here.  Furthermore, XMM-Newton
is able to spatially resolve the Crab, whereas the measurement by SPI and
IBIS/ISGRI may include contributions from the pulsar, possibly leading to
higher fluxes in comparison to the XMM-Newton observations.  Note, that the
difference in flux normalization between XMM-Newton and SPI are beyond the
systematic errors quoted. The model calculations shown here are fixed to the
XMM-Newton spectra which then naturally underpredict the SPI measurements. The
shape of the spectrum measured by SPI has been taken into account as both the
power-law below 100~keV as well as above smoothly connect to the spectra
measured at the low energy end by XMM-Newton and with Comptel at higher
energies.

\begin{figure*}[t!bh]
 \centering 
 \includegraphics[width = 0.9\textwidth]{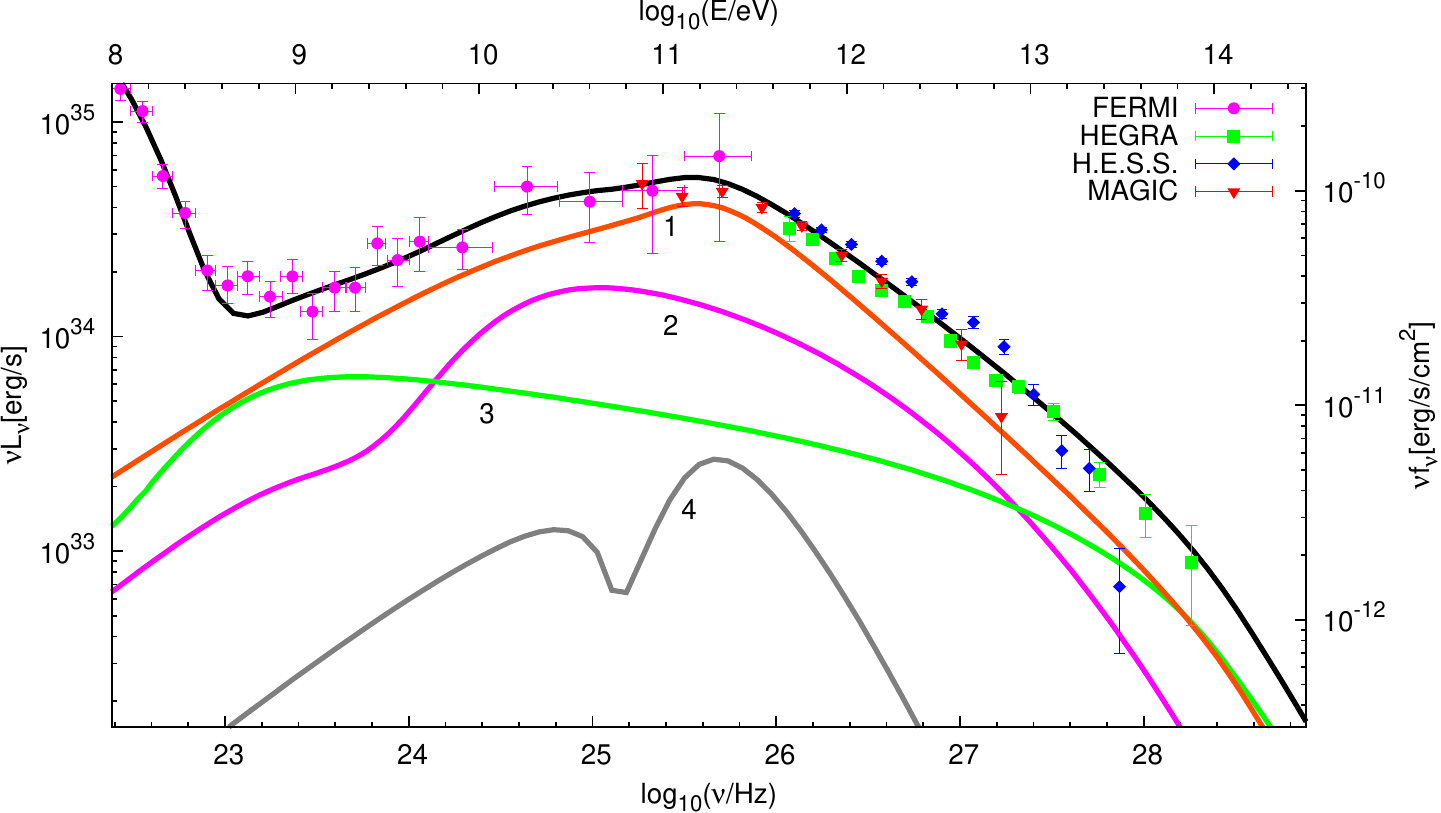}
 \caption{The total IC flux due to different seed photon fields
and the Fermi data points are shown. For the numbering see the
text.}
\label{fig:IC}
\end{figure*}

The predicted IC emission and the Fermi observations are used to determine the
average magnetic field. This is shown in Fig. \ref{fig:IC}. The IC fluxes due
to the different photon fields and electron populations add up to give the
total black solid line. A standard $\chi^2$-minimization is used to determine
the best value for the average $B$-field. Since a varying magnetic field also
changes the synchrotron flux, the underlying electron spectrum is varied
accordingly to compensate for the change, i.e. the synchrotron flux remains
constant. Taking the systematic energy uncertainty on the global energy scale
of the Fermi data into account, $\Delta E/E = {}^{+5\%}_{-10\%}
\,\mathrm{(sys.)}$ \citep[see e.g.][]{2009PhRvL.102r1101A}, the average
$B$-field is found to be \begin{equation} 
\bfield\label{eqn:Bfit}.
\end{equation}

The discussion of this result in the light of MHD 
calculations \citep{1984ApJ...283..694K} can be found in \citet{Meyer}. It is worthwhile noting that the magnetic field derived here is less than half the value of the commonly used $300\,\mu\mathrm{G}$.

\subsection{The electron spectrum of the nebula}
\label{sec:el_spec}

The underlying electron spectrum $\Difft{\Nel}{\gamma}$ is the crucial quantity that determines the shape of the SED \citep[see][for a detailed discussion]{Meyer}. It consists of the two afore mentioned electron populations
where the radio electron spectrum is given by
\begin{eqnarray} 
\Diff{\Nel^r}{\gamma} &=& \left\{\begin{array}{ll} N_0^r \gamma^{-S_r} & \mathrm{for}\quad \gamma^r_1\le\gamma\le\gamma^r_2,\\ \\
0 & \mathrm{otherwise},\end{array}\right.\label{eqn:el_pop}.
\end{eqnarray}
The radio electrons were probably injected in the phase of rapid
spin-down during the initial stages of the pulsar-wind evolution
\citep{1999A&A...346L..49A}. The values for $\gamma_0^r,\gamma_1^r$, and $\gamma_2^r$ 
are summarized in Table \ref{tbl:el_spec}. 
Above $\gamma^r_2$ and
below $\gamma_1^r$ a sharp cut-off for the spectrum is chosen, i.e.
$\Difft{\Nel}{\gamma} = 0$ for $\gamma<\gamma^r_1$ and $\gamma>\gamma^r_2$. 

\begin{table*}[t!hb]
 \centering
 \begin{tabular}{lccccc}
 \centering
 \textbf{Population}	&$ N_0$& $\ln\gamma_0$	& $\ln\gamma_1$	& $\ln\gamma_2$	& $ S$\\
\hline
\hline
Radio		&$120.0(1)$ & -- & $3.1$ & $12.1(7)$ & $1.60(1)$		\\
\hline
Wind		&$78.6(3)$& $19.5(1)$ & $12.96(3)$ & $22.51(3)$ & $3.23(1)$
 \end{tabular}
\caption{Cut-off and normalization energies together with the spectral indices $S$ of the electron spectrum. See text and Eqn. \ref{eqn:wind} for further details.}
\label{tbl:el_spec}
\end{table*}

The wind electrons produce via
synchrotron emission the bulk of the observed SED above sub-mm/FIR
wave-lengths.  The wind electrons are constantly injected downstream 
of the wind shock (hence the name).  The radiatively cooled
spectrum of the wind electron  has a spectral index of
$S_w = 3.23 = 2.23 + 1$ which can naturally be explained by ultra-relativistic
1${}^\mathrm{st}$ order Fermi acceleration with synchrotron cooling \citep[see
e.g.][]{1986MPARp.239.....K}. 
\\
An additional feature present in the hard X-ray spectrum which follows a broken
power-law with a break at $\sim 70\unit{kev}$ requires a break with $\Delta
S=0.43$ in the electron-spectrum. We
tentatively relate this break to the injection mechanism, given that it can
hardly be related to energy dependent escape (the X-ray emitting electrons
suffer cooling well before escaping the nebula). The value of $\Delta S$ could
hint at an energy dependent effect
similar to diffusion in a Kolmogorov-type turbulence power spectrum. 
\\
For high and low
energies the wind electron spectrum cuts off super-exponentially,
\begin{eqnarray}
\Diff{\Nel^w}{\gamma} &=&  N_0^w\left\{\begin{array}{ll}
\left(\frac{\gamma}{\gamma_0^w}\right)^{-S_w},
&\mathrm{for}\quad\gamma<\gamma_0^w, \nonumber\\
\\
\left(\frac{\gamma}{\gamma_0^w}\right)^{-(S_w+\Delta S)},
& \mathrm{for}\quad\gamma^w_0\le\gamma\le\gamma_2^w ,\\
\\
0, & \mathrm{for}\quad \gamma > \gamma^w_2, \\
\end{array}\right\}\nonumber\\
&{}&\quad\times\exp\left(-\left[\frac{\gamma_1^w}{\gamma}\right]^{2.8(4)}\right)\label{eqn:wind}.
\end{eqnarray}
where the values for $\gamma_0^w$,  $\gamma_1^w$ and $\gamma_2^w$ are 
listed in Table \ref{tbl:el_spec}. The total 
energy of the radio and wind electrons, respectively is found to be
\begin{eqnarray}
 E_r & = & mc^2 \int_{1}^{\infty} \,\gamma \Diff{\Nel^r}{\gamma}\,\mathrm{d}\gamma = 3.10\times10^{48}\unit{ergs},\\
 E_w & = & mc^2 \int_{1}^{\infty} \,\gamma \Diff{\Nel^w}{\gamma}\,\mathrm{d}\gamma =2.28\times10^{48}\unit{ergs},
\end{eqnarray}
indicating that the total energy is much smaller than the integrated energy
released through the spin-down of the pulsar.  The fact that both relic
electrons and wind electrons share roughly equal energy is probably
coincidental.

\section{Cross Calibration of IACTs \& Fermi}
\label{sec:crosscal}
The updated model of the SED of the Crab Nebula provides an opportunity for the
cross calibration between ground based air shower experiments and the Fermi
large area telescope. The method is demonstrated here with  the imaging air
Cherenkov telescopes (IACTs) HEGRA, H.E.S.S.  and MAGIC but is applicable to
any other ground based air shower experiment. 

\begin{figure*}[tbh]
 \centering
\subfigure[]{
 \includegraphics[width = 0.45\textwidth]{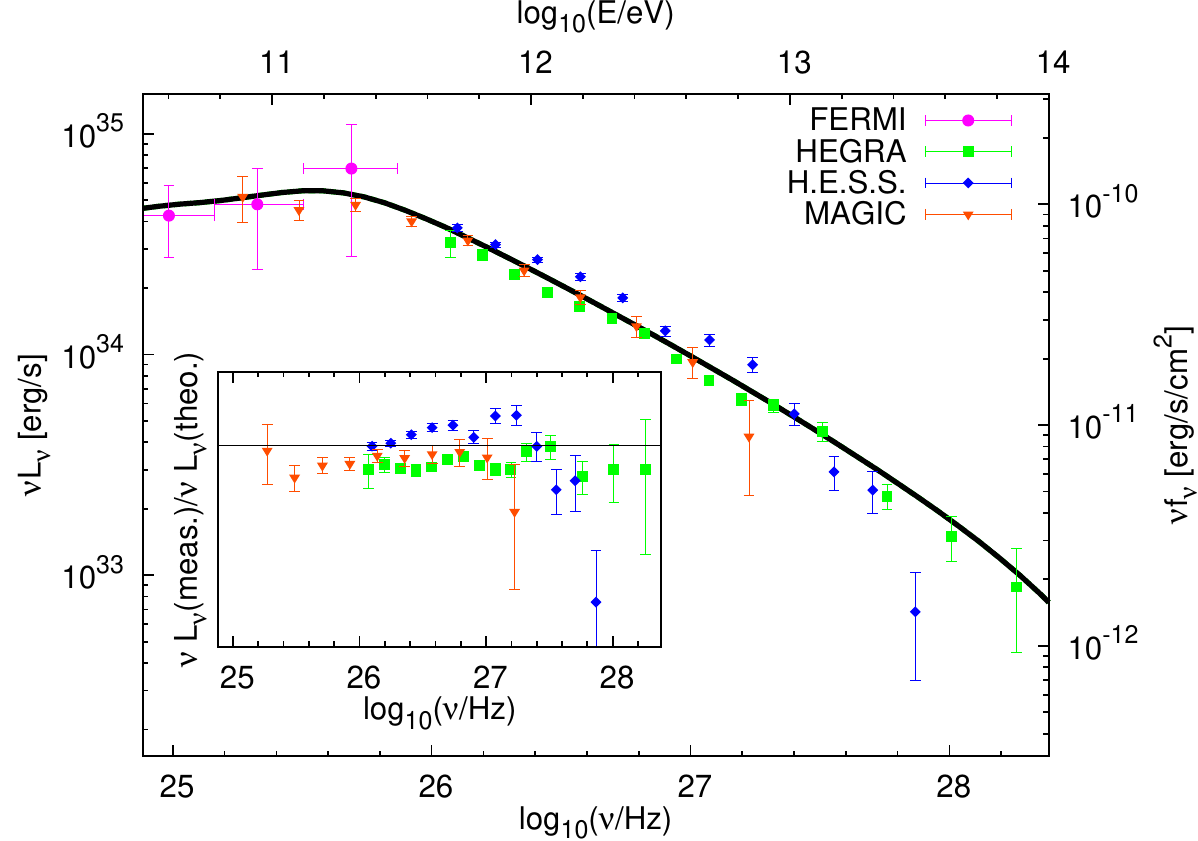}
 \label{fig:unscale}}
 \subfigure[]{
 \includegraphics[width = 0.45\textwidth]{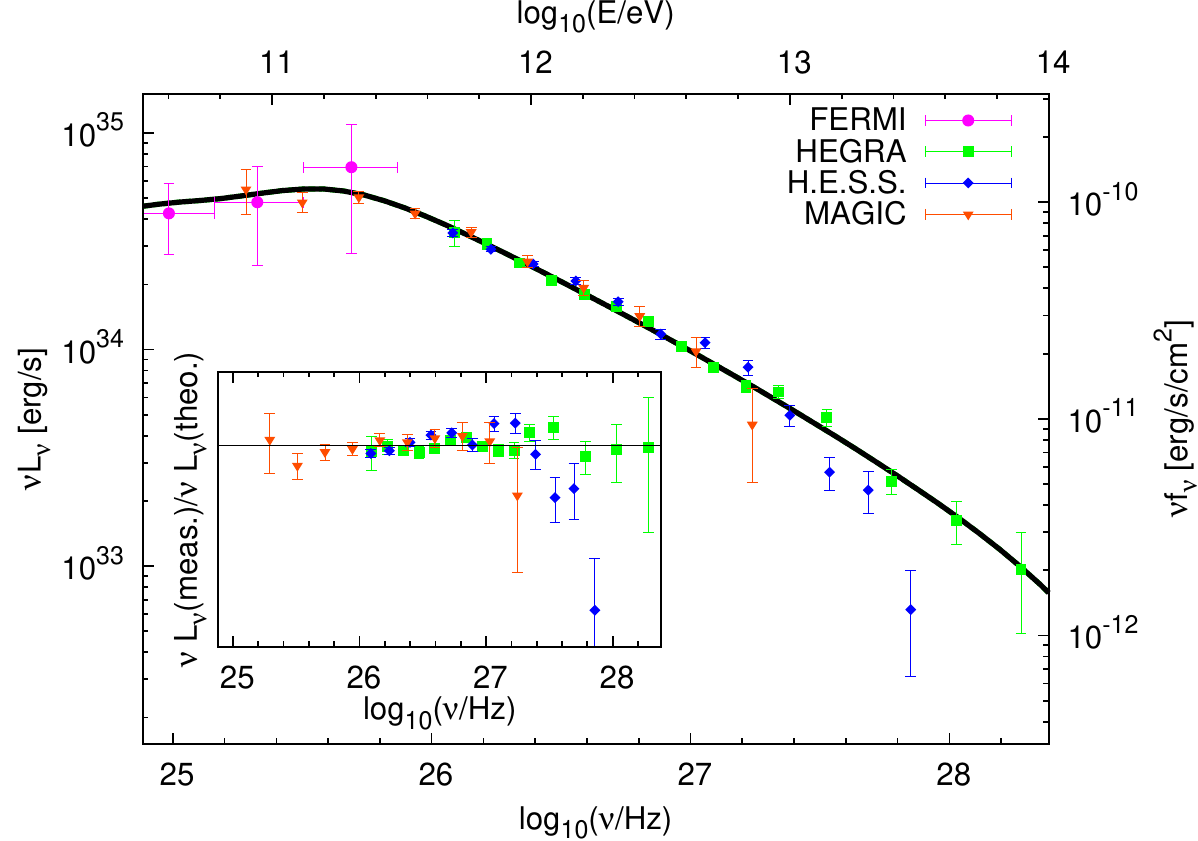}
 \label{fig:scale}}
\caption[]{\subref{fig:unscale} The IC model (solid black line) together with
measurements from IACTs and Fermi. No Enery scaling is applied.
\subref{fig:scale} The same situation as in \ref{fig:unscale} but with the
scaling factors of Eqn. \ref{eqn:scale} and Table \ref{tbl:scale} applied. The open squares denote H.E.S.S. points
that were not included in the fit due to the high systematic uncertainty of the highest energy measurements.}
\end{figure*}

The energy calibration of IACTs is done indirectly with the help of detailed
simulations of air showers and the detector response. However, the remaining
systematic uncertainty on the absolute energy scale of typically 15~\% leads to
substantial differences in the observed flux and position of cut-offs in the
energy spectra between different IACTs and also between Fermi/LAT and IACTs.\\
Since the observed energy spectra are usually quite broad in energy, the
position of features in the spectra are not useful (and may be time-dependent
for some objects) for cross calibration. On the other hand, cross calibration
between Fermi/LAT and IACTs provides indirectly a means of benefitting
from the careful beam-line calibration of the Fermi/LAT \citep[see e.g.][]{2009ApJ...697.1071A}.
 For this reason, the average magnetic field used in the model was fixed to
Fermi observations. \\
The cross calibration is now accomplished in the following
way: for each IACT an energy scaling factor
$s_\mathrm{IACT}$ is introduced such that \begin{equation} E^\prime = E \cdot
s_\mathrm{IACT}.\label{eqn:scale} \end{equation} 
The scaling factor $s_\mathrm{IACT}$ is determined via a
$\chi^2$-minimization in which the energy scale is changed according to
the formula above until the data points reproduce the model best. The scaling
factors for the different instruments are listed in Table \ref{tbl:scale}
together with the statistical errors and the reduced $\chi^2$ values of the
fit. The statistical uncertainties were obtained by summing the errors of the
$\chi^2$-fit and the statistical errors of the model in quadrature. The latter
are mainly due to the uncertainties on the $B$-field of Eqn.
\ref{eqn:Bfit}. To illustrate the result, Figure \ref{fig:unscale} and
\ref{fig:scale} compare the
unscaled data points of the Crab nebula with the scaled ones. 
It is evident that the data points fit the model better after scaling. All scaling factors lie within the afore mentioned 
 15~\% energy uncertainty of the IACTs.

The application of the cross calibration eliminates the systematic uncertainty of the energy scale 
of the IACTs and adjusts the energy scale to the one of the Fermi/LAT. Since the model relies on the Fermi/LAT measurements, 
the Fermi/LAT's absolute energy uncertainty remains. This implies an improvement for the 
systematic uncertainty on the energy scale of IACT measurements from $\pm 15 ~\%$ to ${}^{+5~\%}_{-10~\%}$. 
However, the model itself also contributes to the systematic uncertainties. The main uncertainty stems from the fact that a
constant magnetic field is assumed which is very likely not the case. 
It should be also noted, that the cross calibration factors
hinge mainly on the high statistics measurement of the IACTs at low TeV-energies which are produced by 
electrons that co-exist in the same volume in the nebula as the electrons emitting at lower energies (i.e. in the Fermi/LAT energy range).

\begin{table*}
 \centering
\caption{Energy scaling factors of the IACTs for the cross calibration.}
\label{tbl:scale}
 \centering
 \begin{tabular}{ccccc} 
\textbf{Instrument}	& \textbf{Scaling factor $s_\mathrm{IACT}$}
& \textbf{Stat. error $\Delta s$}	& $\chi^2_\mathrm{before}\slash\mathrm{d.o.f.}$& $\chi^2_\mathrm{after}\slash\mathrm{d.o.f.}$\\
\hline
\hline
Fermi/LAT		& $1$	  	&$+0.05 ~-0.03$	&	--		&$0.49$		\\
HEGRA			& $1.042$ 	&$\pm 0.005$ 	&	$7.652$ 	&$1.046$		\\
H.E.S.S.		& $0.961$ 	&$\pm0.004 $ 	&	$11.84 $	&$6.476$		 \\
MAGIC			& $1.03$ 	&$\pm0.01  $ 	&	$1.671$		&$0.656$
 \end{tabular}
\end{table*}

As a first application of the cross calibration, we   derive upper limits
on the diffuse $\gamma$-ray background.

Both Fermi \citep{2009PhRvL.102r1101A}, and H.E.S.S.
\citep{2008PhRvL.101z1104A,2009arXiv0905.0105H} have measured the cosmic ray $e^- + e^+$ spectrum.
Unlike Fermi, the telescopes from H.E.S.S. cannot accurately distinguish
between showers induced by electrons (or positrons) or photons, such that
up to $\approx 50$~\% of the observed electromagnetic air showers
could be induced by photons. Hence, H.E.S.S.
actually measures electrons and diffuse background photons. Taking the
difference of the two measurements we can derive an upper limit on the 
flux of the $\gamma$-ray background. The scaling factors derived above are now
used to convert the IACT data 
to the same energy scale of Fermi, which reduces substantially the systematic
uncertainty on the observed flux given that the electron spectrum follows a
soft power-law with $E^{-3}$. 

\begin{figure}[tbh]
 \centering
 \includegraphics[width = 0.4\textwidth]{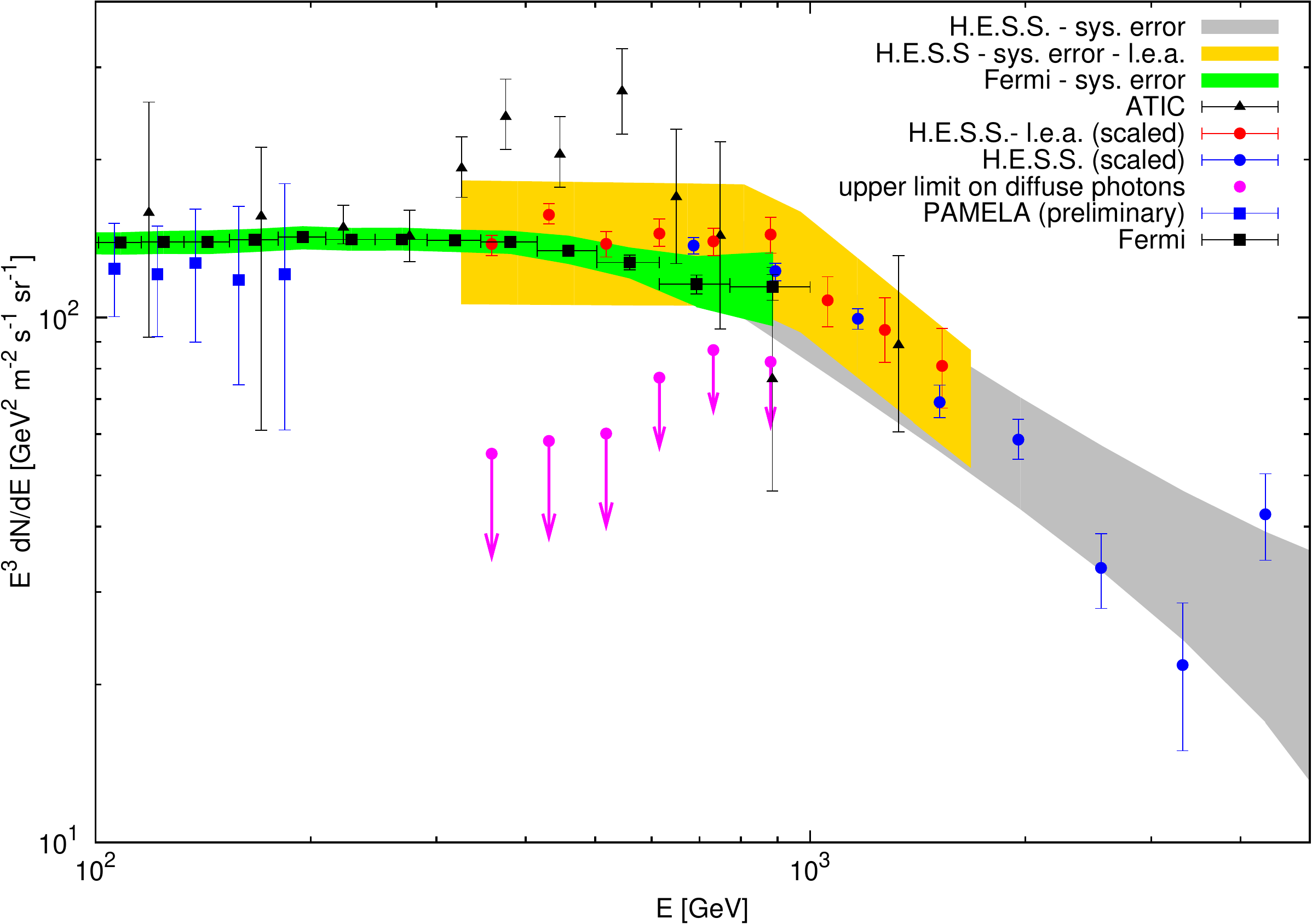}
 \caption{$e^- + e^+$ Spectrum reported by H.E.S.S. and Fermi. The cross calibration is applied, hence the uncertainty on the global energy scale is eliminated.}
 \label{fig:atic}
\end{figure}

The upper limits were derived by subtracting the
two fluxes from the overlapping region of the measurements. This
corresponds to the first six H.E.S.S. points of the low energy analysis in
Figure \ref{fig:atic}. The remaining systematic energy uncertainties, denoted
by the green and yellow bowties, were taken into account for the derivation of
the upper limits: the flux points of the H.E.S.S. measurements were shifted to
their maximum value allowed by the systematic uncertainties while the Fermi
points were shifted to the minimum value. Hence, the result represents a
conservative approximation of the upper limits. 

An important result of the cross calibration is that the peak in the spectrum
observed by ATIC \citep{2008Natur.456..362C} appears more unlikely after
applying the scaling factors.

\section{Summary and Outlook}
\label{sec:summary}

An updated model for the SED of the Crab nebula has been introduced. It
incorporates a new electron spectrum that consists of two electron populations
that have been studied extensively in the past \citep[see
e.g.][]{1996MNRAS.278..525A, 1999A&A...346L..49A}.

The average $B$-field was derived by fixing the IC flux to Fermi/LAT measurements.

The model makes it possible to derive energy scaling factors that eliminate the systematic energy uncertainties of ground based air shower experiments. An application of the cross calibration to the diffuse $\gamma$-background has been presented and upper limits have been derived. Moreover, the excess measured by the ATIC collaboration seems unlikely with the scaled H.E.S.S. observations. 

Nevertheless, an improved model for the emission of the Crab is conceivable. Such a model could comprise a spatially varying $B$-field and hence the fact that electrons emitting different energies are exposed to different field strenghts. 

To conclude, a way to establish the Crab as a \textit{true} standard candle in $\gamma$-ray astronomy has been presented and  versatile applications to other observations of bright steady or pulsed sources (e.g. the Crab Pulsar, the Galactic Center other Pulsar Wind Nebulae, etc.) are imaginable. Moreover, an application of the cross calibration will help to improve dark matter searches and constraints on the extra galactic background light.

\bigskip 
\begin{acknowledgements}
The participation of the Fermi Symposium  was made possible with the support of the German federal ministry for education and research (Bundesministerium f\"ur Bildung und Forschung).

It was also supported by the collaborated research center (SFB) 676 ``Particle, Strings and the early Universe'' at the University of Hamburg.
\end{acknowledgements}

\bigskip 
\bibliographystyle{authordate3}
\bibliography{apj-jour,crab_report_praktikum}

\end{document}